\documentclass[aps,pra,twocolumn,superscriptaddress,shownopacs]{revtex4}

\usepackage{graphicx}
\usepackage{hyperref}
\usepackage{amsmath}
\usepackage{epstopdf}
\usepackage{siunitx}
\usepackage{bm}
\usepackage{ifpdf}
\usepackage{float}
\usepackage{upgreek}
\usepackage{xcolor}
\usepackage{graphicx}
\usepackage{hyperref}
\usepackage{amsmath}
\usepackage{epstopdf}
\usepackage{bbm}

\newcommand{\bra}[1]{\langle #1 |}
\newcommand{\ket}[1]{| #1 \rangle}
\newcommand{\abs}[1]{\left| #1 \right|}

\begin{document}

\title{Connecting heterogeneous quantum networks \\ by hybrid entanglement swapping}

\author{Giovanni Guccione \footnotemark[4]\footnotetext{\footnotemark[4]These two authors contributed equally to the work.}}
\affiliation{Laboratoire Kastler Brossel, Sorbonne Universit\'e, CNRS, ENS-Universit\'e PSL, Coll\`ege de France, 4 Place
Jussieu, 75005 Paris, France}
\author{Tom Darras \footnotemark[4]}
\affiliation{Laboratoire Kastler Brossel, Sorbonne Universit\'e, CNRS, ENS-Universit\'e PSL, Coll\`ege de France, 4 Place
Jussieu, 75005 Paris, France}
\author{Hanna Le Jeannic \footnotemark[3]\footnotetext{\footnotemark[3]Present address: Center for Hybrid Quantum Networks (Hy-Q), Niels Bohr Institute, University of Copenhagen, Blegdamsvej 17, DK-2100 Copenhagen, Denmark.}}
\affiliation{Laboratoire Kastler Brossel, Sorbonne Universit\'e, CNRS, ENS-Universit\'e PSL, Coll\`ege de France, 4 Place
Jussieu, 75005 Paris, France}
\author{Varun~B.~Verma}
\affiliation{National Institute of Standards and Technology, 325 Broadway, Boulder, Colorado 80305, USA}
\author{Sae Woo Nam}
\affiliation{National Institute of Standards and Technology, 325 Broadway, Boulder, Colorado 80305, USA}
\author{Adrien Cavaill\`{e}s}
\email{adrien.cavailles@lip6.fr}
\affiliation{Laboratoire Kastler Brossel, Sorbonne Universit\'e, CNRS, ENS-Universit\'e PSL, Coll\`ege de France, 4 Place
Jussieu, 75005 Paris, France}
\author{Julien Laurat}
\email{julien.laurat@sorbonne-universite.fr}
\affiliation{Laboratoire Kastler Brossel, Sorbonne Universit\'e, CNRS, ENS-Universit\'e PSL, Coll\`ege de France, 4 Place
Jussieu, 75005 Paris, France}


\maketitle

\textbf{Recent advances in quantum technologies are rapidly stimulating the building of quantum networks. With the parallel development of multiple physical platforms and different types of encodings, a challenge for present and future networks is to uphold a heterogeneous structure for full functionality and therefore support modular systems that are not necessarily compatible with one another. Central to this endeavor is the capability to distribute and interconnect optical entangled states relying on different discrete and continuous quantum variables. Here we report an entanglement swapping protocol connecting such entangled states. We generate single-photon entanglement and hybrid entanglement between particle-like and wave-like optical qubits, and then demonstrate the heralded creation of hybrid entanglement at a distance by using a specific Bell-state measurement. This ability opens up the prospect of connecting heterogeneous nodes of a network, with the promise of increased integration and novel functionalities.}

\begin{figure}[b!]
\vspace{-0.4cm}
\includegraphics[width=0.85\columnwidth]{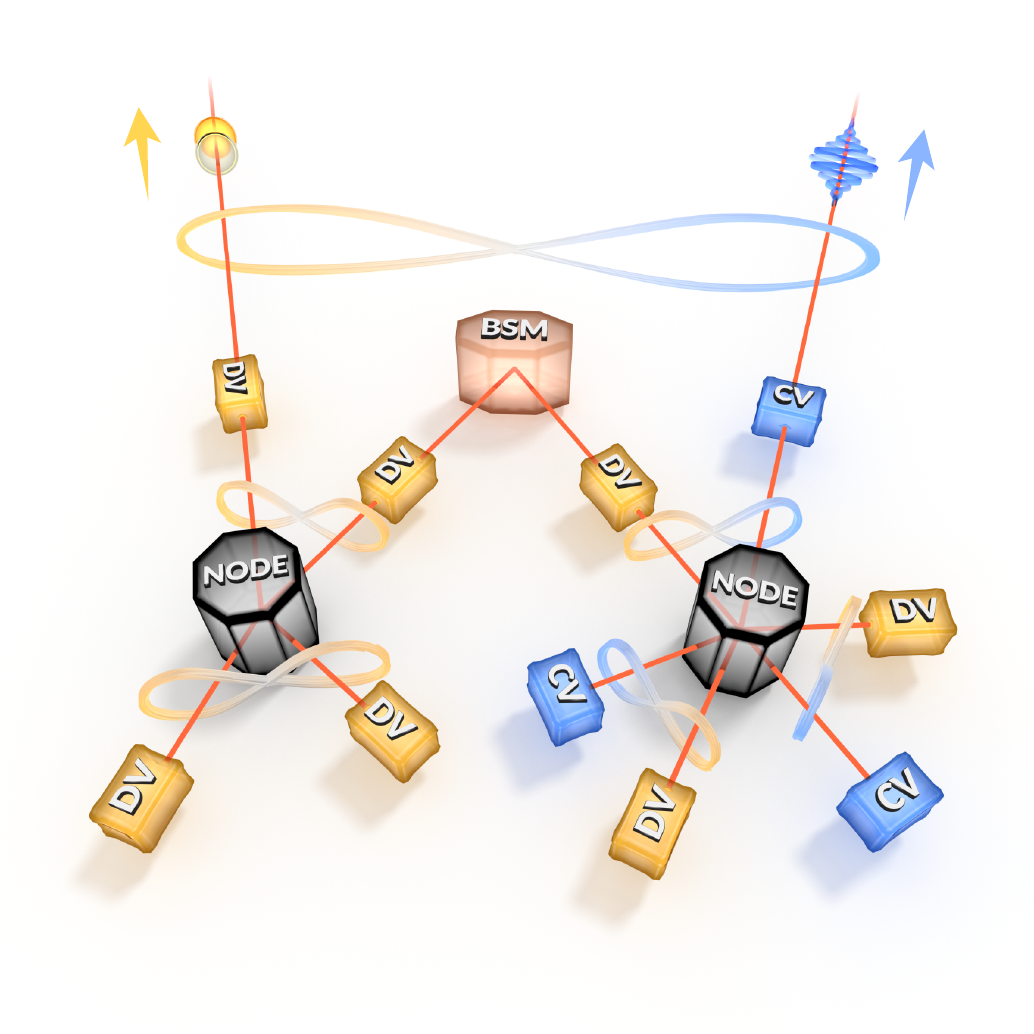}
\vspace{-1cm}
\caption{\textbf{A heterogeneous quantum network linked by entanglement swapping.} A discrete-variable node, on the left, establishes a link to a hybrid node, on the right, via a swapping protocol implemented by a Bell-state measurement (BSM) at an intermediate station. Hybrid CV-DV entanglement is created between two modes that never interacted. The resulting entangled state is available to perform subsequent remote state preparation, teleportation-based encoding conversion or to connect disparate physical platforms at longer distances.}
\label{fig1}
\end{figure}

When looking at compatibility issues in quantum information processing and networks, light is an emblematic example. Its use has historically been split between two communities depending on the degree of freedom favored for encoding. On one side is the continuous-variable (CV) approach, which treats optical fields as waves~\cite{Braunstein:2005:RevModPhys}. On the other side is the discrete-variable (DV) approach, harnessing the properties of individual photons~\cite{Kok2007}. These two strategies have been studied extensively and led to a variety of seminal demonstrations for quantum technologies with complementary advantages ~\cite{OBrien2009,Wamsley2015}. By considering a hybrid approach bridging the two, one could envision a quantum network where the two encodings can be interchanged fittingly to the task at hand. This conversion could also find applications for connecting disparate quantum devices \cite{Kurizki2015}, e.g., CV oscillators and finite-level DV systems, which can couple preferentially to one or the other optical degree of freedom. Achieving such versatility is a critical challenge for developing a modular approach of quantum networks.

\begin{figure*}[t!]
\includegraphics[width=1.77\columnwidth]{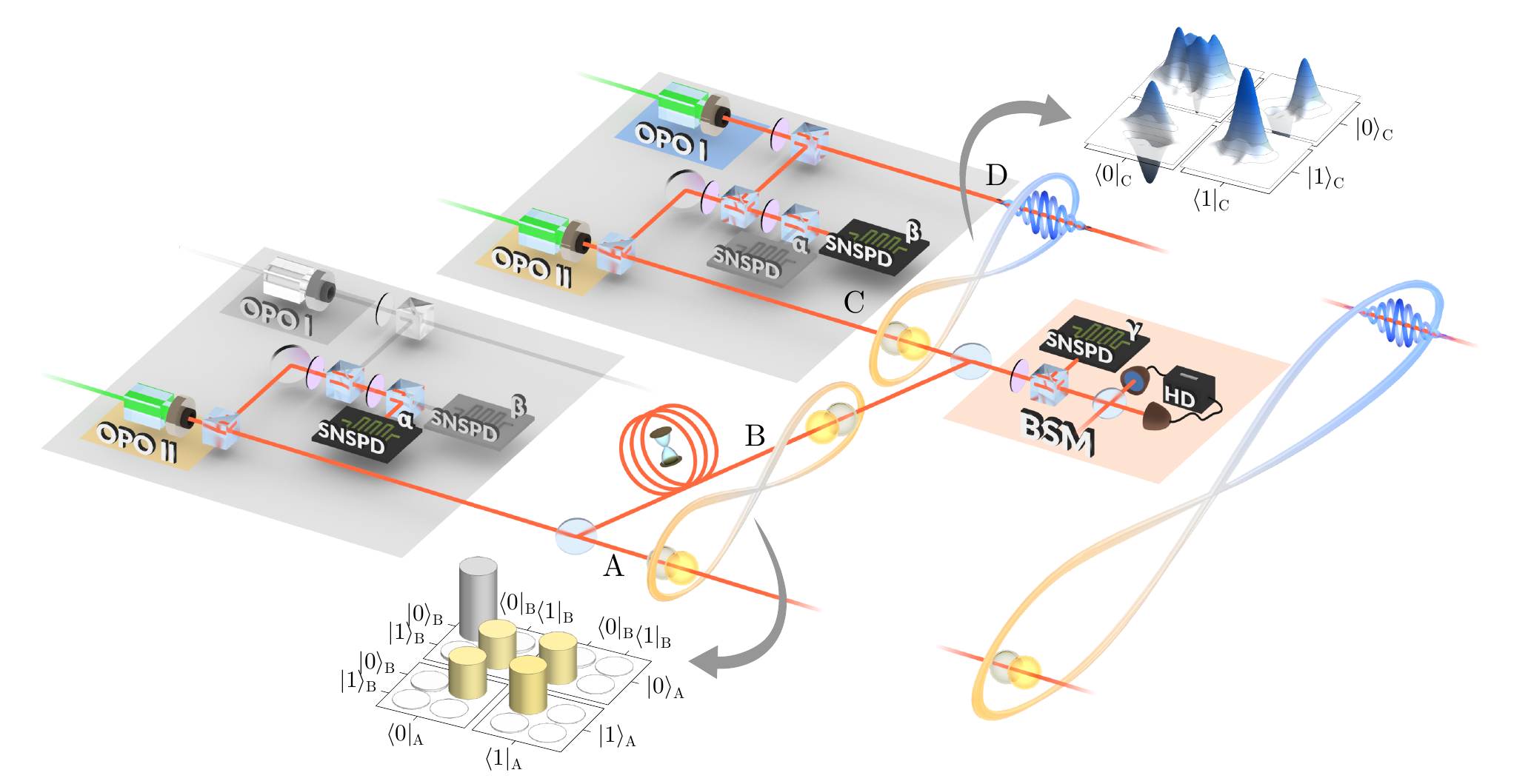}
\caption{\textbf{Experimental setup.} The gray panels outline the non-linear sources and operations, i.e., optical parametric oscillators and heralding measurements. Continuous-variable states are produced via a single-mode squeezer (OPO~I) while discrete-variable states are generated by a two-mode squeezer (OPO~II). The sources are used at different times to generate DV single-photon entanglement between modes $\textrm{A}$ and $\textrm{B}$ first, and hybrid DV-CV entanglement between modes $\textrm{C}$ and $\textrm{D}$ afterwards. The matrices show the measured entangled states: single-photon entanglement heralded by SNSPD~$\alpha$ (bottom), and hybrid entanglement heralded by SNSPD~$\beta$ (top, Wigner representation of the reduced states). To prepare for entanglement swapping, a delay line holds off the state in mode $\textrm{B}$ to enable the temporal matching of one DV mode from each of the two initial states. These modes are mixed at the Bell-state measurement station (BSM, light brown panel), where a combination of a detection event on SNSPD~$\gamma$ and a quadrature measurement via homodyne detection (HD) is performed. Upon success, hybrid entanglement is heralded between modes $\textrm{A}$ and $\textrm{D}$ where high-efficiency homodyne detections are used for full two-mode quantum state tomography.}
\label{fig2}
\end{figure*}

Several experiments have spearheaded the development of the hybrid quantum optics paradigm~\cite{Loock:2011:LaserPhotonRev,Andersen:2015:NatPhys,Minzioni:2019:JOpt}. The first realizations from a decade ago consisted in combining the DV and CV toolboxes to generate non-Gaussian states, such as optical Schr\"odinger cat states starting from Gaussian squeezed light \cite{dakna,grangier,polzik}. These advances spurred intense experimental and theoretical efforts toward novel protocols, including deterministic teleportation of photonic qubits~\cite{Takeda:2013:Nature} or witnesses for single-photon entanglement based on CV homodyne measurements~\cite{Morin2013}. More recently, the engineering of hybrid entanglement of light~\cite{Morin:2014:NatPhot, Jeong:2014:NatPhot, Huang:2019:NewJPhys}, i.e., entanglement between particle- and wave-like optical qubits, enabled remote state preparation~\cite{Le-Jeannic:2018:Optica} and teleportation between different encodings~\cite{Ulanov:2017:PhysRevLett, Sychev:2018:NatComm}. This entanglement was also certified for use in one-sided device-independent protocols~\cite{Cavailles:2018:PhysRevLett}.\\

In this context, a hybrid network requires an original approach to link and transfer entanglement between diverse nodes~\cite{Wehner:2018:Science}. A cornerstone capability is thereby given by entanglement swapping, which is at the foundation of quantum repeaters~\cite{Briegel:1998:PhysRevLett, Duan:2001:Nature}. Entanglement swapping was originally performed for discrete-variable systems~\cite{Pan:1998:PhysRevLett} before being extended to continuous-variable schemes~\cite{Jia:2004:PhysRevLett, Takei:2005:PhysRevLett}. A swapping technique combining the salient characteristic from the two optical paradigms has also been recently demonstrated to transfer DV entanglement via CV entanglement~\cite{Takeda:2015:PhysRevLett}. However, the distribution of hybrid CV-DV entanglement of light by entanglement swapping has not been addressed until now. 

In this work, we report an entanglement swapping protocol where two end nodes arrive at sharing hybrid CV-DV entanglement despite one of the initial stations starting as a discrete-variable-only platform. The scenario is sketched in Fig.~\ref{fig1}, where a DV node establishes a quantum link with a hybrid CV-DV node after swapping is heralded by a specific Bell-state measurement (BSM) at a central station. Specifically, in our experiment we created single-photon entanglement and hybrid entanglement, and then performed swapping using a measurement involving single-photon counter and homodyne detection. Entanglement between the modes that never interacted was finally verified by full quantum state tomography. Our experiment thereby provides a crucial capability for communication in heterogeneous networks. This entanglement distribution has also be shown to be advantageous for optics-based quantum computation \cite{Lee2012}, loss-resilient quantum key distribution~\cite{Sheng:2013:PhysRevA} and entanglement purification~\cite{Bose:1999:PhysRevA, Sheng:2013:PhysRevA}.

The experimental scheme is presented in Fig.~\ref{fig2}. Our implementation is an adaptation of the scenario in Fig.~{\ref{fig1}} where the two input entangled states are generated in sequence using the same resources. More specifically, the continuous- and discrete-variable components of the initial entangled states are respectively derived from single- and two-mode squeezed vacua generated from two optical parametric oscillators by parametric down conversion (OPO~I for single-mode, OPO~II for two-mode). The OPOs are driven well below threshold to ensure high fidelity with the expected states (see Methods). The two-mode squeezer is used at two consecutive times to generate first the single-photon entanglement and then the DV component of the hybrid entangled state. Overall, the experiment amounts to the use of five independent single-mode squeezers. This time-multiplexed usage of the quantum state sources is, however, specific to our implementation and not intrinsic to the swapping protocol hereby presented. The addition of a fast switch would enable the full spatial separation of the modes (see Appendix A).

Specifically, our realization starts with the generation of single-photon entanglement. A polarizing beamsplitter placed at the output of OPO~II separates the signal and idler modes. The idler mode is channelled to a heralding station, where it can be detected after filtering by a high-efficiency superconducting nanowire single-photon detector (SNSPD) ~\cite{Le-Jeannic:2016:OptLett}, with a system detection efficiency of about 85\% and dark noise below 10 counts per second. A first detection event on SNSPD~$\alpha$ heralds the presence of a single photon in the signal mode. The photon propagates to a 50:50 plate where it is split between modes $\textrm{A}$ and $\textrm{B}$, generating the DV entangled state, up to a normalization factor,
\begin{eqnarray}
	\ket{\Phi}_\textrm{AB} = \ket{0}_\textrm{A}\ket{1}_\textrm{B}+\ket{1}_\textrm{A}\ket{0}_\textrm{B}.
	\label{single-photon state}
\end{eqnarray}
Mode $\textrm{B}$ is then directed to a 47-ns delay line in free space (see Appendix A). 

Next, hybrid CV-DV entanglement is created. For this purpose, a small fraction of the single-mode squeezed vacuum ($3\%$) is tapped off and mixed with the idler mode of OPO II. A detection event, now on SNSPD~$\beta$, heralds the entangled state~\cite{Morin:2014:NatPhot}
\begin{eqnarray}
	\ket{\Psi}_\textrm{CD} = \ket{0}_\textrm{C}\ket{\textrm{cat}_-}_\textrm{D}+\ket{1}_\textrm{C}\ket{\textrm{cat}_+}_\textrm{D}.
	\label{hybrid state}
\end{eqnarray}
Here, $\ket{\textrm{cat}_+}$ and $\ket{\textrm{cat}_-}$ denotes the even and odd optical Schr\"odinger cat states~\cite{Minzioni:2019:JOpt} forming the basis of our CV qubit space, corresponding to superpositions of coherent states of amplitude $\alpha\approx0.9$. The single-mode squeezed vacuum and the photon-subtracted squeezed vacuum states generated by the OPO can concurrently reach fidelities above 90\% with these states at moderate squeezing levels of about 5 dB. Details on this measurement-induced preparation have been reported elsewhere~\cite{Morin:2014:NatPhot}.

Given the successful generation of both initial entangled states, with the adequate delay between heralding events (see Appendix B), typically at a rate of 140 events per second, entanglement swapping is then performed. The output of the delay line, mode $\textrm{B}$, and the DV component of the hybrid state, mode $\textrm{C}$, are brought to interfere on a 50:50 beamsplitter, leading to the combined state
\begin{eqnarray}
	\ket{\psi_\textrm{BSM}} & = &\sqrt{2}\ket{0}_\textrm{B}\ket{0}_\textrm{C}\otimes\ket{1}_\textrm{A}\ket{\textrm{cat}_-}_\textrm{D}\nonumber\\
	&&+\ket{0}_\textrm{B}\ket{1}_\textrm{C}\otimes\big(\ket{0}_\textrm{A}\ket{\textrm{cat}_-}_\textrm{D}+\ket{1}_\textrm{A}\ket{\textrm{cat}_+}_\textrm{D}\big) \nonumber	\\
			&&	+\ket{1}_\textrm{B}\ket{0}_\textrm{C}\otimes\big(\ket{0}_\textrm{A}\ket{\textrm{cat}_-}_\textrm{D}-\ket{1}_\textrm{A}\ket{\textrm{cat}_+}_\textrm{D}\big) 	\nonumber\\
			&&	+	\big(\ket{0}_\textrm{B}\ket{2}_\textrm{C}-\ket{2}_\textrm{B}\ket{0}_\textrm{C}\big)\otimes\ket{0}_\textrm{A}\ket{\textrm{cat}_+}_\textrm{D}.	\label{BSM state}
\end{eqnarray}
If one (and exactly one) photon is detected on either of the outputs, modes $\textrm{A}$ and $\textrm{D}$ become hybrid-entangled without ever directly interacting. We consider only detection events on mode $\textrm{C}$ to ensure we always recover the same state $\ket{\Psi}_\textrm{AD}$ at the output. We also note that at every stages of our experiment active phase stabilization of the different paths is a critical requirement (see Methods). 

\begin{figure*}[!t]
\includegraphics[width=1.88\columnwidth]{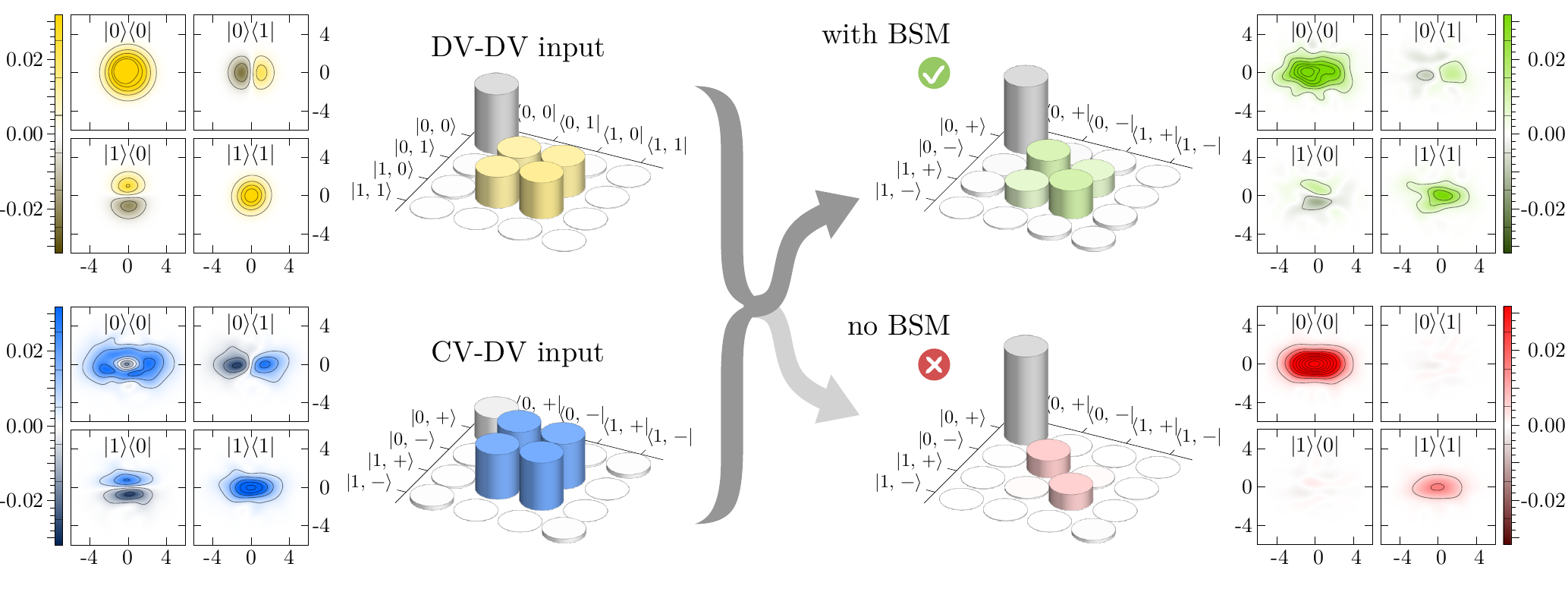}
\caption{\textbf{Entanglement swapping results.} Tomographic measurements of the initial entangled states are given on the left, with DV-DV entanglement on top and hybrid CV-DV entanglement at the bottom. The right side provides the measurement of the output state when the BSM heralds entanglement swapping and, for reference, when no BSM is implemented (see Appendix D for more specific comparison). All states are displayed using two representations, a hybrid density-Wigner plot and a density matrix giving the reconstructed state projected in the relevant basis (vacuum--single-photon $\{\ket{0},\ket{1}\}$ for DV, coherent-state superpositions $\{\ket{\textrm{cat}_+},\ket{\textrm{cat}_-}\}$ for CV). The number of tomographic samples for each state reconstruction is $600000$, $200000$, $7800$ and $200000$ respectively. For simplicity only the real part of the density matrices is shown, as the imaginary part consists of near-zero elements of magnitude smaller than $1\%$. The input states coincide with those featured in Fig.~\ref{fig2}.}
\label{fig3}
\end{figure*}

To realize the single-photon projection, the Bell-state measurement consists of two parts: a low-reflectivity beamsplitter ($R=10\%$) to tap off a small fraction of light, which is sent to a single-photon detector (SNSPD~$\gamma$), and a homodyne setup for quadrature measurement. The role of the SNSPD is to herald a photon subtraction from mode $\textrm{C}$. The homodyne measurement is then used to condition on the quadrature of the state after subtraction to conclude the projection and thereby trigger a successful BSM (see Methods). More details on key parameters are provided in Appendix C. In our specific case, because of the use of a delay line, the beamsplitter for the BSM corresponds to the one used to create the single-photon entangled state, this time accessed from the other input port. This beamsplitter would be distinct in a general scenario involving two separate sources. Overall, the whole procedure inclusive of state preparation and filtering for successful swapping -- i.e., three single-photon detections and one homodyne conditioning -- occurs at a rate of about 3 events per minute.

We now turn to the experimental results. The quantum states involved in our experiment are shown in Fig.~\ref{fig3}. They are measured by quantum state tomography performed with two high-efficiency homodyne detections and reconstructed via maximum-likelihood algorithms~\cite{Lvovsky}. 

We first show, on the left-hand side, the two initial heterogeneous entangled inputs: single-photon DV-DV entanglement (top) and hybrid CV-DV entanglement (bottom). To characterize the DV-DV input we used homodyne tomography on mode $\textrm{A}$ directly and on mode $\textrm{B}$ after the delay line. The DV-DV state shown in Fig.~\ref{fig3} is corrected only for homodyne detection loss, corresponding to $18\%$ for both modes. Similarly, the hybrid CV-DV input has been reconstructed using homodyne tomography on modes $\textrm{C}$ (without the delay line's beamsplitter) and $\textrm{D}$. In this case the detection loss amounts to $17\%$ for the DV mode and $15\%$ for the CV mode. For the two initial states, with these measurements in hand, we finally computed the negativity of entanglement~\cite{Vidal} given by $\mathcal{N}=\left(||\rho^{T_X}||_1-1\right)/2$, where $\rho^{T_X}$ stands for the partial transpose of the two-mode density matrix $\rho_{XY}$ with respect to mode $X$. This quantity reaches 0.5 for an ideal maximally entangled state. We obtained $0.099\pm0.002$ for single-photon entanglement and $0.223\pm0.003$ for hybrid entanglement. These values are set by the initial purity of the generated states, slightly degraded relative to previous works from our group \cite{Morin:2014:NatPhot,Le-Jeannic:2016:OptLett} due to a larger pumping power that was necessary to increase the operating rates and also owing to the optical losses in the complex circuit that involves the delay loop.

Next, on the right-hand side of Fig.~\ref{fig3}, we provide the output state between modes $\textrm{A}$ and $\textrm{D}$. As the homodyne setup used for tomography of the DV mode is also the one employed for the BSM, an additional $10\%$ loss due to the low-reflectivity beamsplitter used for photon subtraction is added. A total loss of $26\%$ on the DV mode and $15\%$ on the CV mode is corrected for the reconstruction. The output state at the top corresponds to the outcome when the BSM is successfully implemented. Given this swapping heralding event, non-zero off-diagonal coherence terms appear, thereby confirming the emergence of quantum correlations even though the modes have never directly interacted. We further assessed the success of the swapping protocol by computing the negativity of entanglement, estimated at $0.044\pm0.009$ (see Appendix E). Finally, we repeated the experiment in the same conditions but without the Bell-state measurement. As shown by the state at the bottom, the output modes are not correlated and no entanglement can be observed in this case. These results show the ability to swap between disparate optical entangled states and to establish hybrid CV-DV entanglement between heterogeneous nodes.

\begin{figure}[!t]
\includegraphics[width=0.9\columnwidth]{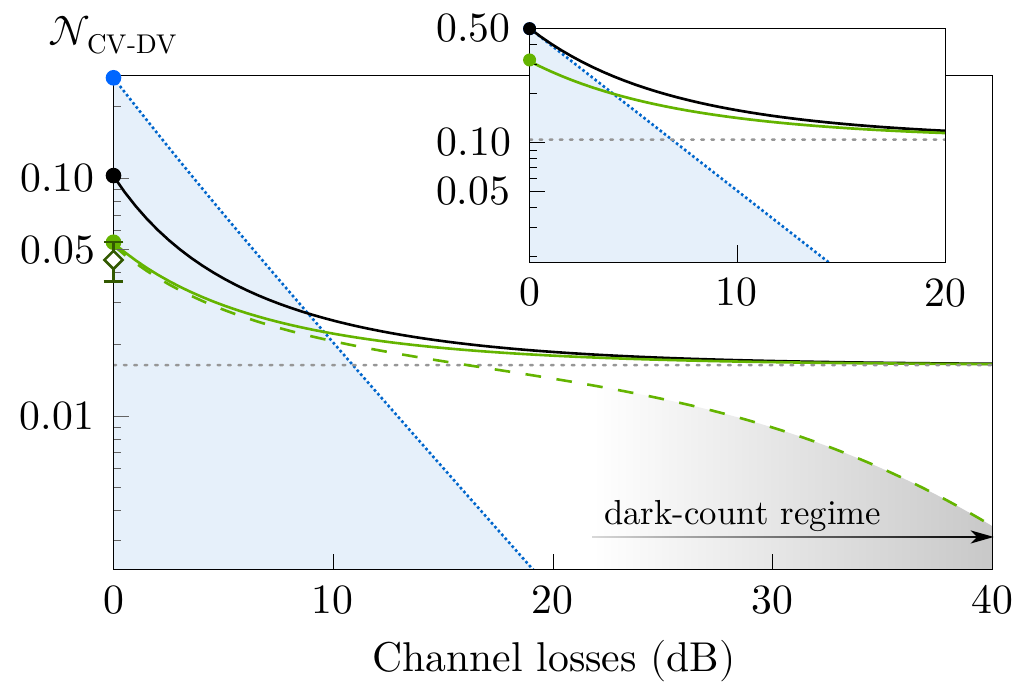}
\caption{\textbf{Swapping for remote hybrid entanglement distribution.} The expected entanglement negativity after swapping of the experimental initial resources is evaluated as a function of the channel loss (considered symmetric on the two channels). The black solid line corresponds to an ideal BSM with vanishing reflectivity $R$ and quadrature conditioning window $\Delta$, while the green line corresponds to the implemented case ($R=10\%$ and $\Delta$ equal to the vacuum shot noise). The dotted grey line corresponds to a BSM without homodyne conditioning. The effect of dark counts is presented in dashed green ($1\%$ of total events). The white point gives the measured output negativity, in good agreement with the simulation. As a reference, the blue line shows the negativity for direct propagation. The inset provides the same plots but starting from maximally-entangled input states and without dark counts.}
\label{fig4}
\end{figure}

The swapping process is of paramount importance in the context of quantum connection as it allows the distribution of entanglement over nodes that would be too distant for direct propagation. Thus, in Fig. \ref{fig4}, we provide theoretical predictions of the negativity of entanglement as a function of transmission losses. In particular, starting from our initial experimental states or from maximally entangled states (inset), we compute the negativity of entanglement of the output hybrid state under different BSM implementations and compare to direct propagation in the most resilient case of symmetric losses on the two transmission channels. The white point indicates our measurement, in good agreement with the simulations. 

As can be seen, the swapping protocol would beat the direct propagation for losses over $4$~dB (about $20$~km of fiber at telecom wavelength) for maximally-entangled input states, and over $9$~dB (about $45$~km) in our experimental conditions. We also note that homodyne conditioning leads to an increase in entanglement compared to a partial BSM (i.e., performed with only single-photon detection and no homodyne conditioning). This is most significant over shorter distances and in the case of highly entangled input states where the two-photon component after mixing is larger. Ultimately, any practical implementation of the protocol will be limited by the heralding rate of the input states and its ratio with the dark-count rate of the used detectors. In our experiment, where the ratio is approximately 1:100, we expect a significant decrease of negativity over $20~\rm{dB}$ of channel losses, as shown in Fig. \ref{fig4}. Additional consequences linked to the state reconstruction as well as false-positive events are detailed in the Appendix F. These results confirm the protocol's performance and its demonstration over longer distances could be achieved in future implementations, although with increasingly challenging rates. 

For practical operations, the protocol's success rates will indeed need to be greatly improved. The current limits are primarily due to the probabilistic nature of the generation process. Different paths are being developed to solve this general bottleneck in quantum technology when cascaded operations are performed. Better synchronization of heralded probabilistic resources, as implemented in recent works with high state purity \cite{Yoshikawa2013,Bouillard2019}, or extension of quantum state engineering and of the hybrid approach to on-demand non-Gaussian sources \cite{Loredo2019, Hacker2019} are promising directions. Implementations at telecom wavelength, or with dual frequencies, would also be possible given the current developments of efficient squeezers in this regime \cite{Mehmet}. 

In conclusion, our work introduced and demonstrated an entanglement swapping protocol that connects nodes with different optical encodings. This core capability opens attractive opportunities for the establishment of remote hybrid quantum links and the development of heterogeneous quantum networks, enabling more versatile quantum information interconnects. An exciting future prospect will be to couple such hybrid entanglement to matter systems and to link thereby not only different encodings but also quantum devices of a different nature with complementary functionalities.\\ 

\begin{center} 
\textbf{Methods}
\end{center}

\noindent{\bf Quantum state engineering}\\
The triply resonant OPOs are based on broadband ($65$~MHz) semi-monolithic linear cavities, pumped below threshold by a frequency-doubled Nd:YAG laser (Innolight GmbH). For both cavities, the input mirror is directly coated on one face of the non-linear crystal, with high reflectivity at $1064$~nm and $95\%$ reflectivity at the $532$~nm pump wavelength. The output mirrors have a radius of curvature of $38$~mm and a coating with $90\%$ reflectivity at $1064$~nm and high reflectivity at $532$~nm. For this experiment OPO~I ($10$-mm type-I phase-matched PPKTP Raicol Crystals) was pumped by $15$~mW of $532$-nm light ($80\%$ below threshold, squeezing of about 5 dB), while OPO~II ($10$-mm type-II KTP Raicol Crystals) was pumped by $5$ mW ($95\%$ below threshold to limit the multiphoton components). The hybrid entangled state is produced by combining the tapped beam from OPO~I and the idler beam from OPO~II and projecting them onto the same mode with a polarizing beamsplitter. The state, heralded on SNSPD~$\beta$, is maximally entangled when the polarizations before the beamsplitter are rotated to have equal counts from each OPO~\cite{Huang:2019:NewJPhys}. This balancing condition places a constrain on the relative counts on the other port of the beamsplitter. With $N_\textrm{I}$ ($N_\textrm{II}$) being the count rates from OPO~I (II), a detector on the unbalanced port would see events at a rate of $\abs{N_\textrm{II}^{2}-N_\textrm{I}^{2}}/(N_\textrm{I}^2+N_\textrm{II}^{2})$. In our case this corresponds to 92$\%$ of photons coming from OPO~II. This strong imbalance enables the heralding of a single photon used for discrete-variable entanglement on SNSPD~$\alpha$. Before detection on the SNSPDs, the non-degenerate modes of the OPOs are filtered out using a $0.5$-nm bandwidth interferential filter (Barr Associates) and a home-made Fabry-P\'erot cavity (free spectral range $330$~GHz, bandwidth $320$~MHz).\\

\noindent{\bf Experimental lockings and phase stability}\\
The two OPO cavities are locked on resonance using the Pound-Drever-Hall technique ($12$-MHz phase modulation on the pump). Seed beams at $1064$~nm, originating from a triangular tilt-locked mode-cleaner cavity (length $40$~cm), are used for phase calibrations. They are first phase-locked with each pump using micro-controllers (ADuC7020 Analog Devices, $1\%$ phase noise). The relative phase between the conditioning paths from the OPOs~I and II, as well as the filtering cavities, are digitally locked ($3\%$ phase noise) using the same technique. The free-space delay line is digitally locked (Newport LB1005 Servo Controller) on a Fano-like resonance shape obtained by slightly tilting the polarization of the beam before self-interference, in a technique similar to the H\"ansch-Couillaud scheme ($3\%$ phase noise). The seed beams are also used for phase calibration of the homodyne detections. Apart from the mode cleaner and OPO cavities, all locks are performed under a sampling-and-hold cycle: with the SNSPDs shut off, the locks are readjusted during a $50$-ms time window thanks to the $1064$-nm seed beams; then, the SNSPDs paths are reactivated to acquire the data during the following $50$~ms while the seed beams are off.\\

\noindent{\bf The Bell-state measurement}\\
The BSM consists of two operations: photon subtraction and quadrature conditioning. The subtraction is implemented by a low-reflectivity beamsplitter. The probability of subtracting a single photon scales linearly with the reflectivity, whereas the probability of a two-photon subtraction scales quadratically. It is thus important to operate in the limit of small reflectivity. Once a single-photon subtraction is heralded, the $\ket{1}_\textrm{C}$ element in the state $\ket{\psi_\textrm{BSM}}$ (Eq.~\ref{BSM state}) reduces to a vacuum state, $\ket{0}_\textrm{C}$. The homodyne measurement is used to favor this vacuum term over other unwanted contributions, i.e., the former two-photon term reduced to a single-photon contribution after subtraction. Having different marginal distributions, the two states have different probability of returning a given quadrature value, meaning that they can be discriminated if the outcome is conditioned to be around a specific point where the states would be orthogonal. Specifically, the probability of detecting a vanishing quadrature value is high for vacuum and negligible for single photon. In our experiment we use a beamsplitter reflection of $10\%$ and a homodyne conditioning window spanning half the standard deviation of the vacuum shot noise on either side of the origin (with a probability overlap of $8\%$ between the two states to be discriminated). The requirements for both of these parameters can be relaxed or strengthened to trade between fidelity and success rate (see Appendix C). It should be noted that in our case where the two-photon component is negligible the efficiency of the SNSPD does not affect the fidelity of the swapped state, but only the success rate. To the contrary, the efficiency of the homodyne detection is crucial for the fidelity. This problem is mitigated by the availability of photodiodes with near-unity efficiency.\\

\noindent \textbf{Acknowledgements:} We thank O.\ Morin and K.\ Huang for their contributions in the early stages of the experiment. \textbf{Funding:} This work was supported by the European Research Council (Starting Grant HybridNet), the PERSU program from Sorbonne Universit\'e (ANR-11-IDEX-004-02), and the French National Research Agency (HyLight project ANR-17-CE30-0006). V.B.V.\ and S.W.N.\ acknowledge funding for detector development from the Defense Advanced Research Projects Agency (DARPA) Information in a Photon and QUINESS programs. G.G. was supported by the European Union (Marie Curie Fellowship HELIOS IF-749213) and T.D. by Region Ile-de-France in the framework of DIM SIRTEQ. \textbf{Author contributions:} G.G.\ and T.D.\ contributed equally to the work. G.G., T.D., and A.C.\ performed the experiment, developed implementation techniques and realized the data analysis. H.L.J.\ contributed to the preparation of the setup. V.B.V.\ and S.W.N.\ developed the single-photon detectors. J.L.\ designed research and supervised the project. All authors discussed the results and contributed to the writing of the manuscript.

\renewcommand{\thefigure}{A\arabic{figure}}

\cleardoublepage

\onecolumngrid

\section*{APPENDIX A: EXPERIMENTAL SEQUENCE}

\begin{figure}[b!]
\begin{minipage}[b]{0.35\linewidth}
\centering
\caption{\textbf{Experimental setup and heralding events.} One time-multiplexed SNSPD ($\alpha/\gamma$) is used sequentially to herald the single-photon entanglement and the BSM. Hybrid entanglement is heralded by SNSPD $\beta$. The acousto-optic modulator (AOM) is switched on by the first detection on SNSPD~$\alpha/\gamma$ to avoid subsequent events from the same heralding path. The free-space delay line is experimentally realized by an optical loop of delay $\Delta t_\textrm{DL}$. Heralding of the BSM is deferred by a fiber delay line until a time $t=\Delta t_\textrm{DL}+\Delta t_\textrm{fiber}$ to allow complete switching off of the AOM.}
\label{figtimemultiplexing}
\end{minipage}
\hspace{0.1cm}
\begin{minipage}[b]{0.6\linewidth}
\centering
\includegraphics[width=1\columnwidth]{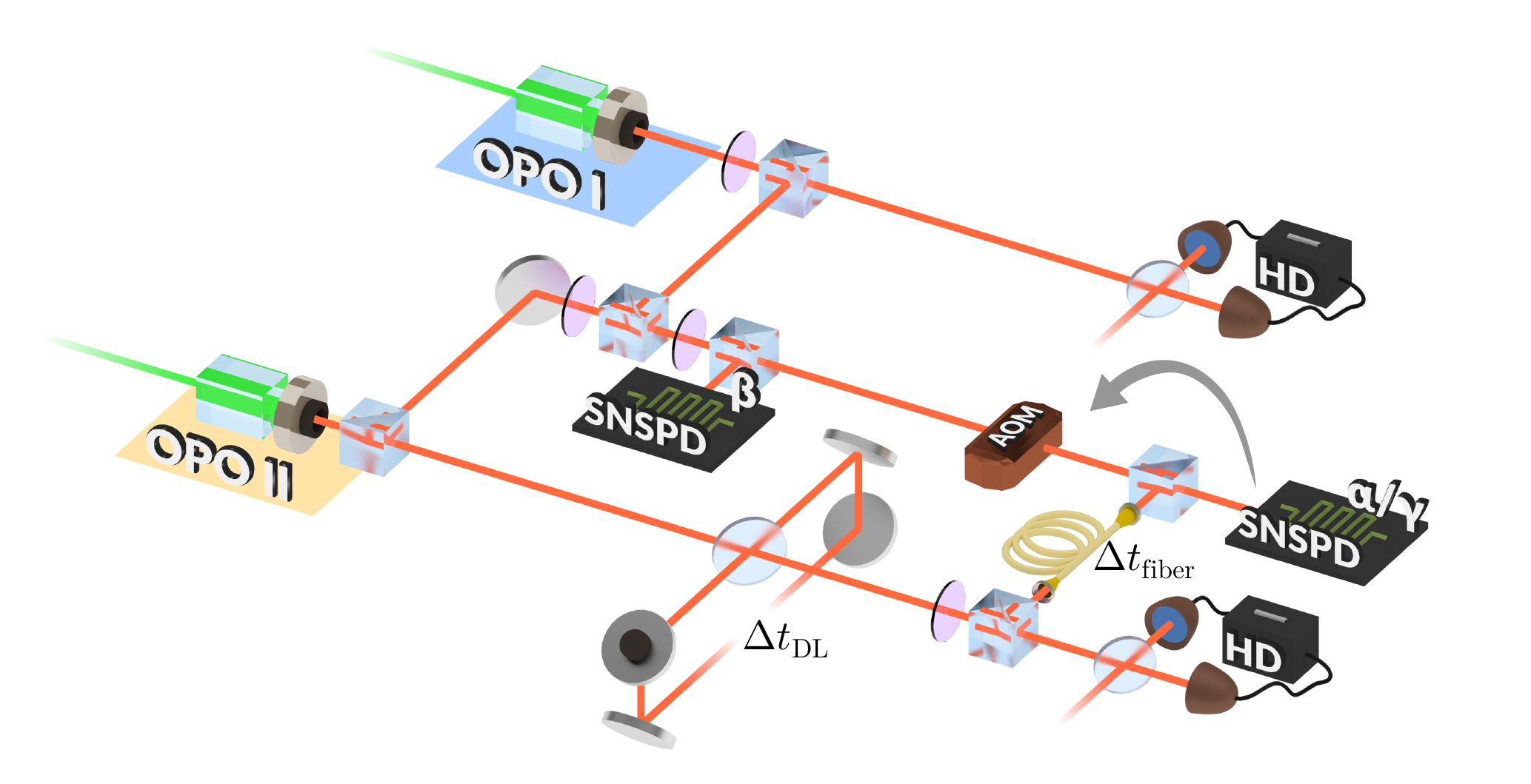}
\end{minipage}
\end{figure}

The OPOs are used sequentially at two different times to generate the single-photon entanglement and the hybrid CV-DV entanglement. This time separation is enabled by a $14$-m free-space loop delay line with a single-trip retardation time $\Delta t_\textrm{DL} = 47$~ns, as shown in Fig. \ref{figtimemultiplexing}. The delay line ($15\%$ transmission loss) is realized by a series of spaced mirrors (Thorlabs NB1-K14). First, a detection event at time $t=0$ on SNSPD~$\alpha$ heralds the generation of a single photon at the output of OPO~II. The temporal mode of the photon, given by $f(t) = \sqrt{\pi\gamma} e^{-\pi\gamma\abs{t}}$ where $\gamma$ is the OPO's full-width half-maximum~(39), has a temporal width $\tau\approx 20$~ns. The state is sent to a 50:50 beamsplitter in order to generate single-photon entanglement. One mode of this entangled state propagates directly to homodyne detection, while the other propagates in the free-space delay line. The travel time inside the delay line is long enough to enable the temporal discrimination between successive DV modes. At $t=\Delta t_\textrm{DL}$, we wait for a coincidence detection event on SNSPD~$\beta$ to herald a hybrid CV-DV entangled state. The DV mode of the hybrid state interferes on the same 50:50 beamsplitter with the delayed mode of the single-photon entangled state.

We then herald the BSM with SNSPD~$\gamma$, which in our case coincides with SNSPD~$\alpha$ used to herald the discrete-variable entanglement. To this end we use a polarizing beamsplitter to have the relevant modes co-propagate with orthogonal polarizations before the detector. In order to avoid double events from the single-photon heralding path, the corresponding beam is shut off with an acousto-optic modulator (AOM) in single-path configuration ($80\%$ extinction of the zeroth order) triggered by the first detection event (SNSPD~$\alpha$). We use a $160$-m fiber delay line (Thorlabs SM980-5.8-125, $\Delta t_\textrm{fiber}=750$~ns) to ensure the temporal distinction of the two events on the same detector (dead time of $100$~ns) and the efficient shutting of the beam ($600$~ns).

Overall, the whole sequence proceeds thereby as follows: first an event on SNSPD~$\alpha/\gamma$ at $t=0$, then an event on SNSPD~$\beta$ at $t=\Delta t_\textrm{DL}$, finally an event on SNSPD~$\alpha/\gamma$ combined with quadrature conditioning on homodyne detection at $t=\Delta t_\textrm{DL} + \Delta t_\textrm{fiber}$. The effective success rate of the experiment in this multiplexed configuration is~$\sim0.05$~Hz and it would increase up to a few hertz by using three independent detectors.

We note that, due to the looped configuration of the delay line, the same homodyne detection is used for the BSM and for the characterization of the swapped state. The swapped DV mode cannot thereby propagate freely in our proof-of-principle demonstration. To overcome this feature, one would need either a longer delay or an optical switch operating faster than $\Delta t_\textrm{DL}$ placed before the detection setup in order to deliver the propagating mode in time. Even though this goal goes beyond the scope of our setup, it is feasible using existing technology~(40). Such an experimental scheme would allow the implementation of complex experiments using minimal resources multiplexed in time. Here, using only two OPOs and two SNSPDs, we achieved a five-mode experiment conditioned on three-fold coincidences.

\section*{APPENDIX B: IDENTIFICATION OF THE EVENTS}

\begin{figure}[b!]
\begin{minipage}[b]{0.35\linewidth}
\centering
\caption{\textbf{Histograms of the detection events on the two SNSPDs.} The BSM is set at $t=0$ and prior events up to $900$~ns earlier are recorded on SNSPD~$\alpha$ for the heralded single-photon and on SNSPD~$\beta$ for hybrid entanglement respectively. The triggering condition on SNSPD~$\gamma$ realigns the occurrences according to the BSM event (temporally detected last). The dashed lines represent the time window of $8$~ns chosen for time filtering (close-up in the inset panels). (a) Events from SNSPD~$\alpha$ (triggering events at $t\geq0$ not shown for clarity). (b) Events from SNSPD~$\beta$.}
\label{figtimefilteringI}
\vspace{-0.6cm}
\end{minipage}
\hspace{0.1cm}
\begin{minipage}[b]{0.6\linewidth}
\centering
\includegraphics[width=1\columnwidth]{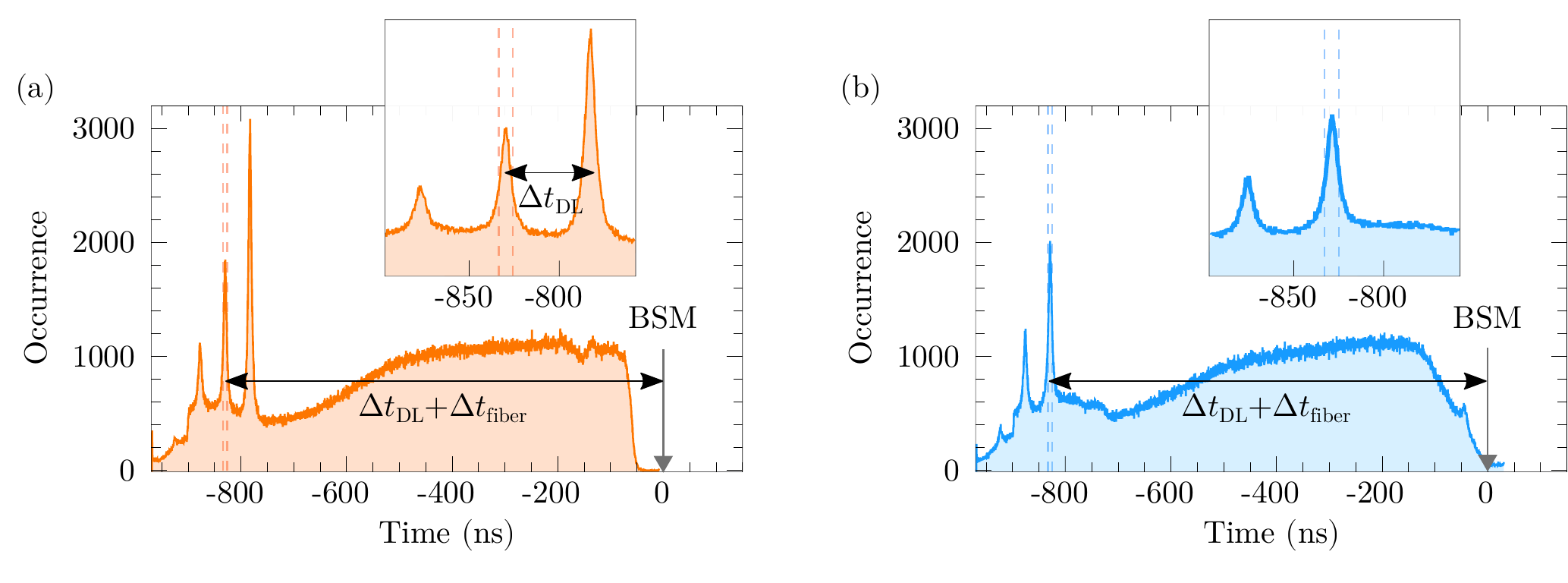}
\end{minipage}
\end{figure}

The data acquisition is performed by accumulating three-fold coincidences in a temporal window of $900$~ns. This is done by having the acquisition trigger coincide with the BSM event at $t=0$ and recording prior events occurring up to $900$~ns earlier on SNSPD~$\alpha$ and SNSPD~$\beta$. We can then select the triggers corresponding to the correct events in time. Figure \ref{figtimefilteringI} provides the histograms of the detection events on both SNSPD~$\alpha$ and SNSPD~$\beta$. Peaks at specific times appear, shaped by the temporal mode of the photons and whose occurrence is linked to the delay line.

In Fig.~\ref{figtimefilteringI}(a) three peaks can be identified. The first peak from the right, at time $t=-\Delta t_\textrm{fiber}$, originates from heralded single photons that propagated straight through the 50:50 beamsplitter and did not enter the free-space delay line. The second peak, at $t=-\Delta t_\textrm{DL}-\Delta t_\textrm{fiber}$, corresponds to photons that have been delayed once, which are thereby the ones of interest for our protocol. At $t=-2\Delta t_\textrm{DL}-\Delta t_\textrm{fiber}$ a smaller peak comes from photons that circulated twice in the delay line. Because the coincidence bias is tied to the triggering event, i.e., the heralding of the BSM, we can see multiple peaks on the histogram of events from SNSPD~$\beta$ in Fig.~\ref{figtimefilteringI}(b) as well. These peaks are again echoes corresponding to multiple passes in the delay line. This time, however, the event relevant to our experiment to herald hybrid entanglement is given by the first, direct set of events at $t=-\Delta t_\textrm{DL}-\Delta t_\textrm{fiber}$. 

The noise floor on these histograms corresponds to independent, stochastic coincidence events that carry no quantum correlation and thus induce vacuum contribution in our measured swapped state. The noise level is suppressed in correspondence of the peaks thanks to the AOM used to shut off the DV heralding path about $600$~ns after the detection of the first event. The resulting signal-to-noise ratio improvement was key to our measurement. Such problem would not persist should three independent SNSPDs be used.

\begin{figure}[t!]

\begin{minipage}[b]{0.35\linewidth}
\centering
\caption{\textbf{Variance scans of the homodyne signals.} (a-b) Variance of the (a) CV and (b) DV homodyne signals as a function of time, before time filtering. The plots are based on a set of $2500000$ recorded events. Peak $\#1$ corresponds to the hybrid entanglement heralded at $t=0$ when the BSM is unsuccessful. In (b), successive peaks are due to different echoes from the delay line. (c-d) Variance of the (c) CV and (d) DV homodyne signals, after time filtering. The higher noise comes from the smaller statistical sampling, which is reduced to 3300 after filtering. The complete disappearance of peak $\#1$ (and corresponding echoes) indicates an efficient filtering. The Bell-state measurement is completed by conditioning on quadrature values around $0$ in correspondance of peak $\#2$ in (f), i.e., when the DV-DV and CV-DV entangled states are mixed. The CV and DV quadratures of the swapped state are obtained by integration of the homodyne signal around peak $\#2$ in (c) and $\#3$ in (d).}
	\label{figtimefilteringII}
\end{minipage}
\hspace{0.1cm}
\begin{minipage}[b]{0.6\linewidth}
\centering
\includegraphics[width=1\columnwidth]{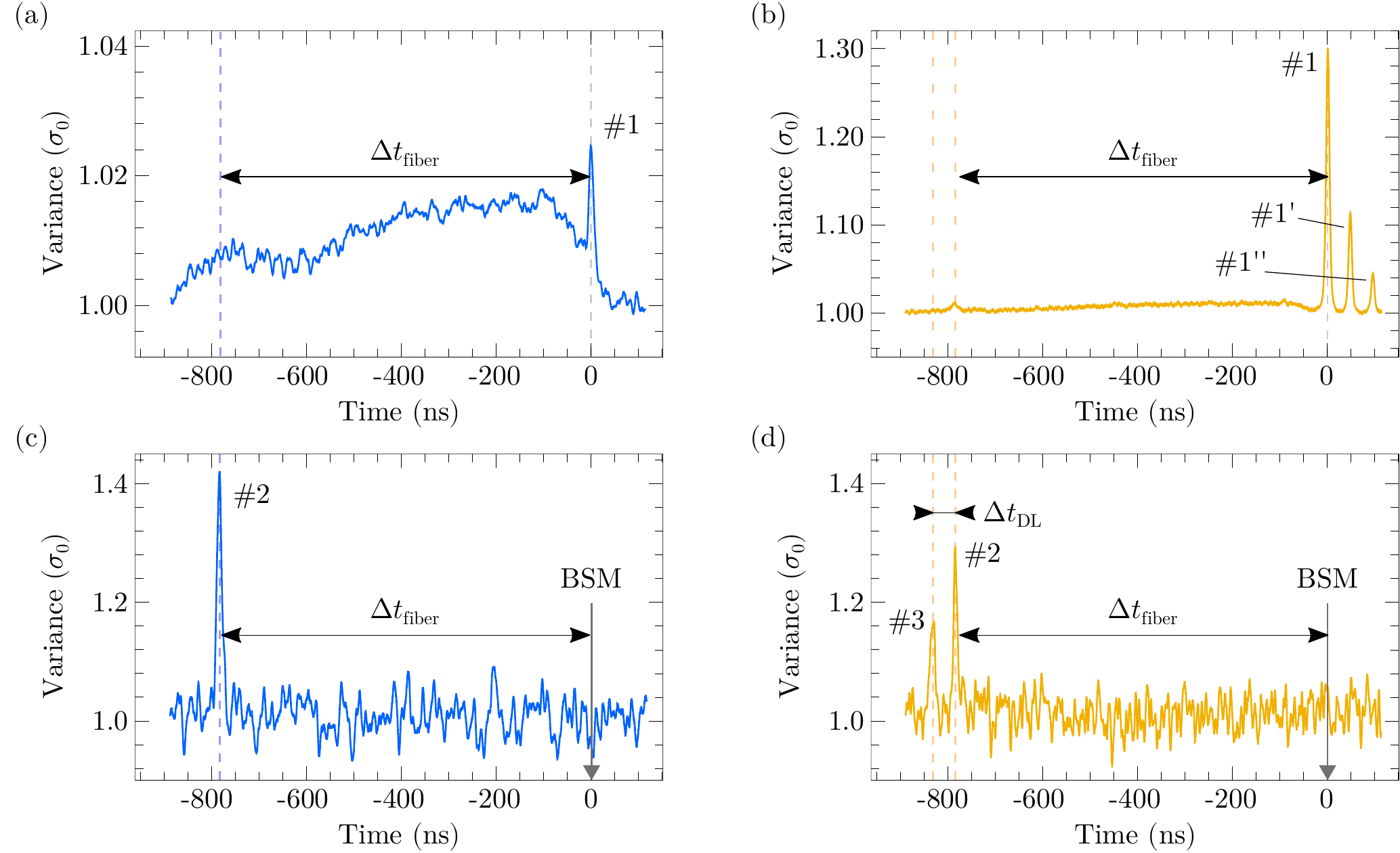}
\vspace{0.6cm}
\end{minipage}
\end{figure}

From all these acquired coincidences we select only those occurring in a temporal window of $4$~ns around $t=-\Delta t_\textrm{DL}-\Delta t_\textrm{fiber}$ (dashed lines in Fig.~\ref{figtimefilteringI}). For the specific data run showed, this filtering preserves $\approx900$ coincidence events. The effect of time selection can be directly seen on the homodyne signals. First, figure \ref{figtimefilteringII}(a) and \ref{figtimefilteringII}(b) give the variance of the homodyne signals during the $900$~ns temporal window for the output CV and DV modes, calculated over the full data set (i.e., without filtering for the right coincidences). Higher values of variance are expected for both CV and DV modes in correspondence of photon-subtracted or single-photon states, respectively. The variance of the CV mode in Fig.~\ref{figtimefilteringII}(a) exhibits a peak (\#1 at $t=0$) given by the hybrid entangled state which is observed in correspondence of the BSM when the coincidence trigger is not the correct one for swapping. The same applies to the other events in the same trace, only that in those cases they are not temporally aligned to the trigger and they result in spread-out noise above the normalization level. The variance of the DV mode in Fig.~\ref{figtimefilteringII}(b) exhibits pronounced peaks (\#1 at $t=0$, \#1' at $t=\Delta t_\textrm{DL}$ and \#1'' at $t=2\Delta t_\textrm{DL}$) corresponding to the heralded single photon and its echoes through the free-space delay line. A faint noise slightly above the shot-noise level, similar to the one of Fig.~\ref{figtimefilteringII}(a), is also noticeable. Figure ~\ref{figtimefilteringII}(c) and \ref{figtimefilteringII}(d) then provide the variances when the time filtering for the right coincidence events is applied ($900$ remaining coincidence events). One can clearly see the disappearance of the peaks \#1, \#1' and \#1'', which indicates efficient filtering since no hybrid or single-photon state should be expected at that time. Even more significantly, what was previously hidden under the noise now appears as temporal-mode-shaped peaks (\#2 and \#3) at $t=-\Delta t_\textrm{fiber}$ and $t=-\Delta t_\textrm{DL}-\Delta t_\textrm{fiber}$. The first peak in temporal order, \#3, corresponds to the heralded single photon from which the DV-DV entanglement is generated. The second peak, \#2, is given by the coincidence of the two entangled inputs, DV-DV and hybrid CV-DV. The BSM homodyne conditioning is performed at this time, $t=-\Delta t_\textrm{fiber}$, on the DV mode (Fig.~\ref{figtimefilteringII}(d)). The two-mode output state obtained after swapping is measured in correspondence of peak \#2 on the CV mode and peak \#3 on the DV mode.

After time filtering and homodyne quadrature conditioning, the homodyne modes of the swapped state are processed through quantum state tomography. By applying a phase propagation to realign different runs together, we were able to combine nine sets of data and reconstruct the entangled state from a total of $7800$ filtered events.

\section*{APPENDIX C: PERFORMANCE OF THE BELL MEASUREMENT}

Our Bell-state measurement consists of two operations: photon subtraction and quadrature conditioning. Thanks to a  beamsplitter with low reflectivity $R\ll1$ for the first operation, we have that the probability of subtracting a single photon scales linearly as $\mathcal{O}(R)$, whereas the probability of a two-photon subtraction scales quadratically and is of order $\mathcal{O}(R^2)$. The regime of low reflectivity is therefore important to avoid two-photon events, even though this condition affects the overall efficiency of the protocol. In our experiment we use a beamsplitter reflection $R=10\%$. Recalling the expression for the combined state before the BSM,

\begin{eqnarray}
	\ket{\psi_\textrm{BSM}} &\propto&\sqrt{2}\ket{0}_\textrm{B}\ket{0}_\textrm{C}\otimes\ket{1}_\textrm{A}\ket{\textrm{cat}_-}_\textrm{D}+\nonumber\\
	&&\ket{0}_\textrm{B}\ket{1}_\textrm{C}\otimes\big(\ket{0}_\textrm{A}\ket{\textrm{cat}_-}_\textrm{D}+\ket{1}_\textrm{A}\ket{\textrm{cat}_+}_\textrm{D}\big) +\nonumber	\\
			&&	\ket{1}_\textrm{B}\ket{0}_\textrm{C}\otimes\big(\ket{0}_\textrm{A}\ket{\textrm{cat}_-}_\textrm{D}-\ket{1}_\textrm{A}\ket{\textrm{cat}_+}_\textrm{D}\big) +	\nonumber\\
			&&		\big(\ket{0}_\textrm{B}\ket{2}_\textrm{C}-\ket{2}_\textrm{B}\ket{0}_\textrm{C}\big)\otimes\ket{0}_\textrm{A}\ket{\textrm{cat}_+}_\textrm{D},	\label{BSM state}
\end{eqnarray}
we see that the heralding of a subtraction event on mode $\textrm{C}$ reduces the first term, corresponding to $\ket{1}_\textrm{C}$, to a vacuum state, $\ket{0}_\textrm{C}$. Similarly, the two-photon component reduces to a single-photon contribution. The quadrature marginal distribution of a vacuum state is maximum around zero, while for single-photon states it vanishes. Therefore, the two states can be discriminated by conditioning for quadrature measurements around the origin. Using the standard deviation of the vacuum shot noise $\sigma_0$ as a scale, in our experiment we consider a homodyne conditioning window $\Delta$ spanning $0.5~\sigma_0$ on either side of the origin, with a probability overlap of $8\%$ between the two states to be discriminated. 

The overall balance of the BSM is determined by the two key parameters, $R$ and $\Delta$. For both, the requirements can be strengthened or relaxed to trade between fidelity and success rate. In principle the BSM is equivalent to the ideal projection $\ket{1}\bra{1}$ when both $R$ and $\Delta$ tend to zero, as the first corresponds to a better approximation of an ideal photon subtraction and the second reduces the overlap of the marginal distributions. However these limiting cases would lead to no events and the swapping would not be heralded. In the following we simulate the BSM to investigate its functional dependence on these parameters. In particular, we study how they affect the count-rate efficiency, the quantum fidelity between the output state and the targeted hybrid entanglement ($\ket{\Psi} = \ket{0}\ket{\textrm{cat}_-}+\ket{1}\ket{\textrm{cat}_+}$) obtained after a $\ket{1}\bra{1}$ projection, and the purity of the final state.

\begin{figure}[t!]
\begin{minipage}[b]{0.35\linewidth}
\centering
\caption{\textbf{Performance of the BSM as a function of the photon-subtracting beamsplitter reflectivity $R$.} (a) Count-rate efficiency for a conditioning window $\Delta=0$ (orange) and $\Delta=\sigma_0$ (red). (b) Fidelity between the output state and the ideal hybrid state $\ket{\Psi}$, again for $\Delta=0$ (cyan) and $\Delta=\sigma_0$ (blue). Dashed lines correspond to the model accounting for losses of $30\%$ on overall single-photon detection and $15\%$ on homodyne detection. $R=10\%$ corresponds to the experimental value.}
		\label{figBSMI}
\end{minipage}
\hspace{0.1cm}
\begin{minipage}[b]{0.6\linewidth}
\centering
\includegraphics[width=\columnwidth]{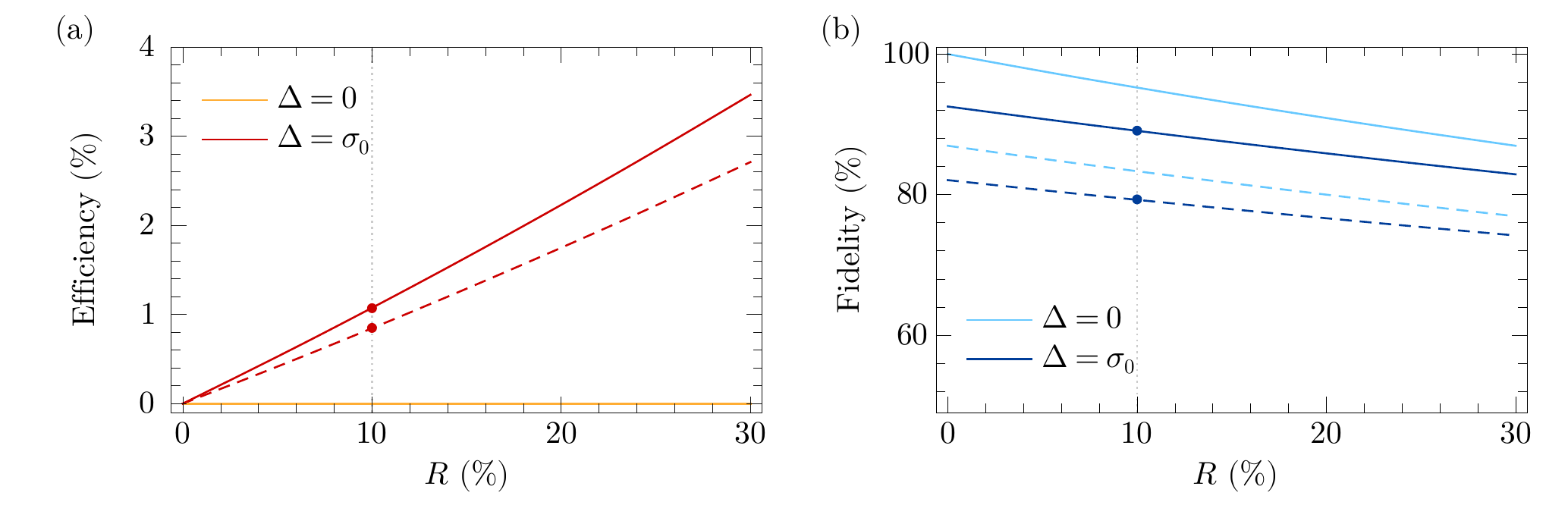}
\vspace{0cm}
\end{minipage}
\end{figure}

\begin{figure}[t!]

\begin{minipage}[b]{0.35\linewidth}
\centering
\caption{\textbf{Performance of the BSM as a function of the width of the conditioning window $\Delta$.} (a) Count-rate efficiency for photon-subtracting beamsplitter reflectivity $R=0$ (orange) and $R=10\%$ (red). (b) Fidelity between the output state and the ideal hybrid state $\ket{\Psi}$, again for $R=0$ (cyan) and $R=10\%$ (blue). Dashed lines correspond to the model accounting for losses of $30\%$ on overall single-photon detection and $15\%$ on homodyne detection. The value $\Delta=\sigma_0$ used in the experiment is marked in both panels.}
	\label{figBSMII}
\end{minipage}
\hspace{0.1cm}
\begin{minipage}[b]{0.6\linewidth}
\centering
\includegraphics[width=\columnwidth]{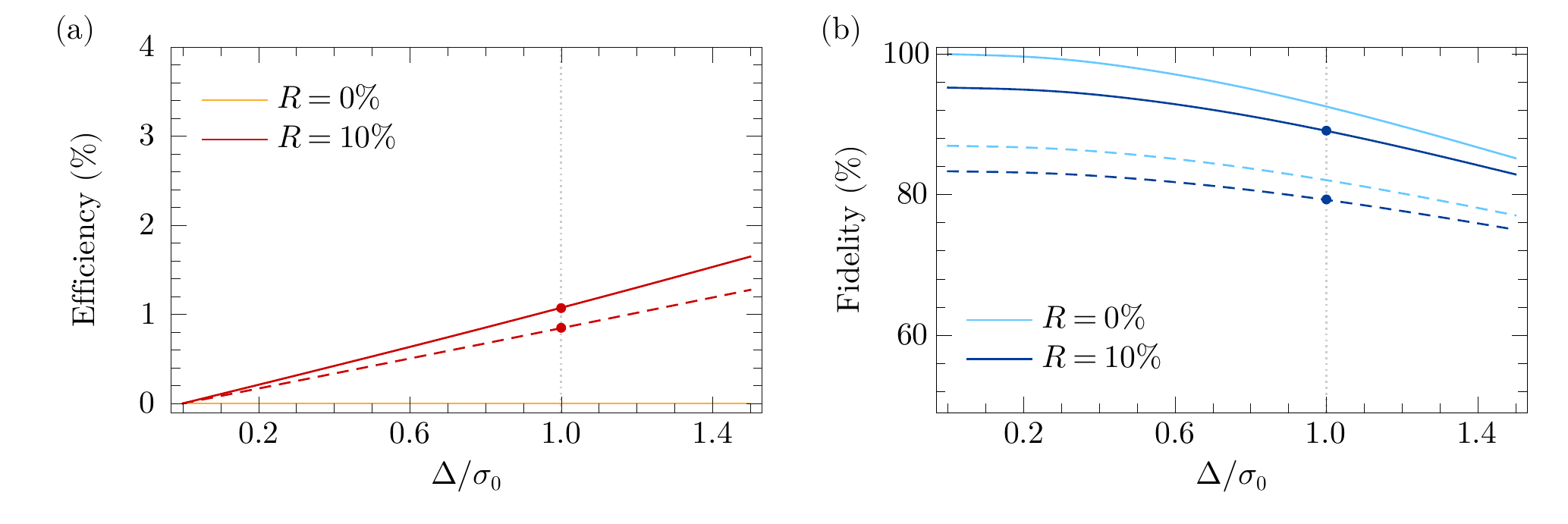}
\vspace{0.cm}
\end{minipage}
\end{figure}

Figure \ref{figBSMI} first provides the count-rate efficiency and the state fidelity as a function of the reflectivity $R$. When $\Delta\to0$, the count-rate efficiency (Fig.~\ref{figBSMI}(a), orange line) is zero independently of the value of $R$. In this limit, however, the fidelity of the output state (Fig.~\ref{figBSMI}(b), cyan line) goes to its ideal value of $1$ for low $R$ values. This is because, for small reflectivity, the probability of subtracting one photon scales linearly while the probability for more than one photon scales at least quadratically in $R$. With a finite value of $\Delta$ the count-rate efficiency increases. We consider the width $\Delta=\sigma_0$ used in the experiment, where $\sigma_0$ is the standard deviation of the vacuum shot noise. In these conditions the count-rate efficiency (Fig.~\ref{figBSMI}(a), red line) scales linearly with $R$, and is on the order of $1\%$ when $R=10\%$. This efficiency gain comes with a decrease in fidelity (Fig.~\ref{figBSMI}(b), blue line), which is expected to be lower than $5\%$ overall, and by $3.5\%$ specifically at $R=10\%$. We can also account for finite efficiencies of the detectors and consider overall single-photon detection loss of $30\%$ and homodyne detection loss of $15\%$ (dashed lines). While the count-rate efficiency remains similar, the expected fidelity reduces noticeably. The lower fidelity of the state is primarily due to the misjudgment of lost photons as qualifying vacuum events and does not depend on the SNSPD loss -- it can be seen that, in fact, the drop in fidelity is effectively a direct translation of the homodyne loss.

\begin{figure}[t!]
\begin{minipage}[b]{0.35\linewidth}
\centering
\caption{\textbf{Purity of the output state}. The purity is given as a function of (a) $R$, (b) $\Delta$, and (c) both. The gray lines correspond to the limiting cases ($\Delta, R\to 0$), the black traces show instead the expected outcome for the chosen experimental conditions ($\Delta=\sigma_0$, $R=10\%$). Dashed lines correspond to the model accounting for losses of $30\%$ on overall single-photon detection and $15\%$ on homodyne detection. The highlighted points correspond to our experimental parameters.}
	\label{figBSMIII}
\end{minipage}
\hspace{0.1cm}
\begin{minipage}[b]{0.6\linewidth}
\centering
\includegraphics[width=\columnwidth]{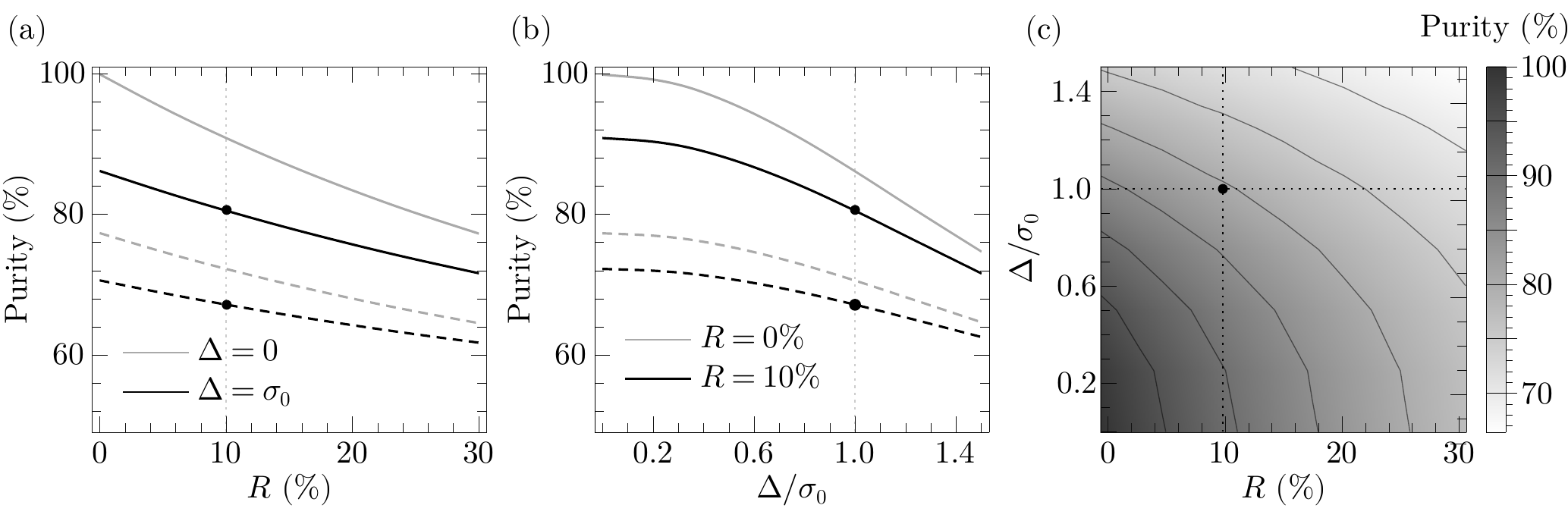}
\vspace{-0.3cm}
\end{minipage}
\end{figure}

Similar trends are observed when keeping the reflectivity of the photon-subtracting beamsplitter fixed and varying the width of the homodyne conditioning window, as shown in Fig.~\ref{figBSMII}. For small reflectivity, the count-rate efficiency increases linearly with $\Delta$. While more events are allowed by the larger window, the overlap of the vacuum's and single-photon's marginal distributions also increases and the fidelity of the output state is reduced accordingly. The reduction is however less pronounced compared to the one observed for increasing $R$, allowing a bit more flexibility in the choice of the working parameter regime. Finally, we analyse in Fig.~\ref{figBSMIII} how the purity of the output state degrades with increasing $R$ and $\Delta$. For the parameters chosen in the experiment the purity is not expected to exceed $\approx 80\%$.\\

\begin{figure}[b!]
\includegraphics[width=0.85\columnwidth]{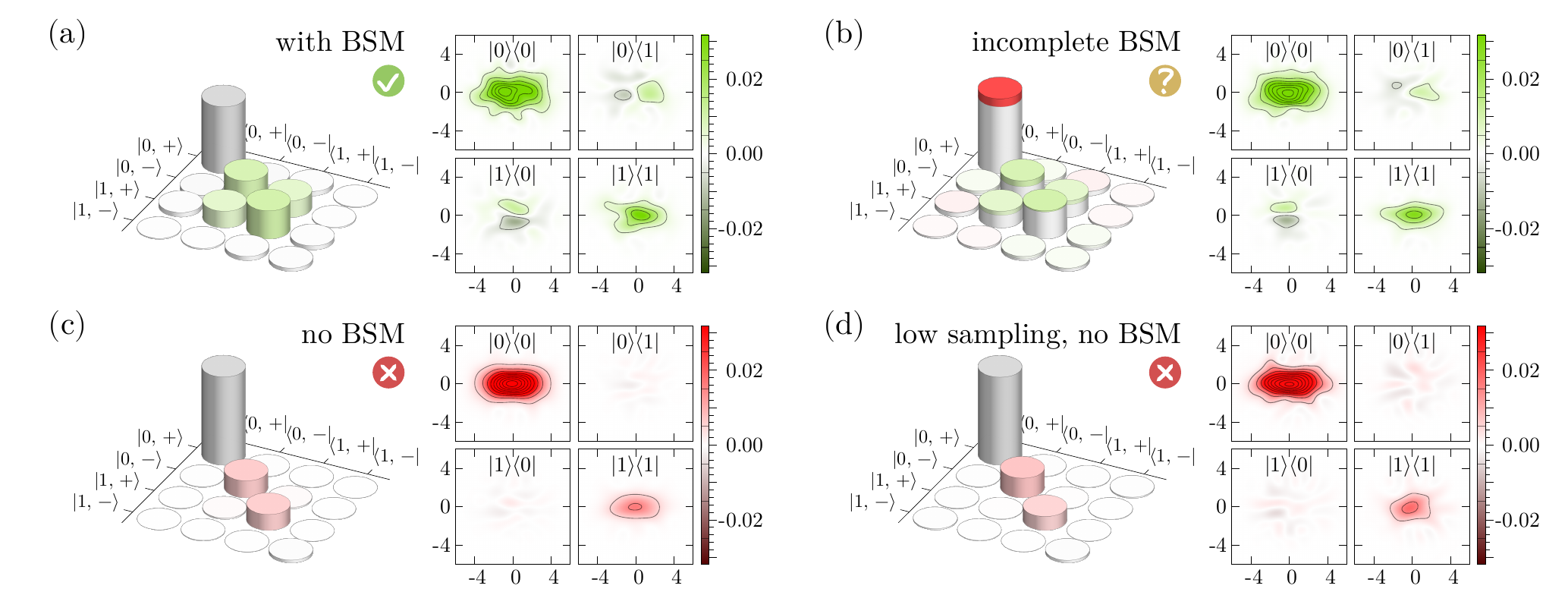}
\caption{\textbf{Comparison of the output state in different configurations.} (a) Output state following the full BSM. This is the result presented in the main text, repeated here for comparison. (b) Output state following a partial BSM, i.e. without homodyne quadrature conditioning. (c) Output state without any BSM. (d) Output state without any BSM, reconstructed from a downsampled data set to match the main result. The number of tomographic samples for each state reconstruction is respectively $7800$, $26500$, $200000$ and $7800$ respectively. In all panels: the left-hand side represents the projection of the reconstructed two-mode density matrix in the $\{\ket{0, \textrm{cat}_+}, \ket{0, \textrm{cat}_-}, \ket{1, \textrm{cat}_+}, \ket{1, \textrm{cat}_-}\}$ basis; the right-hand side shows the hybrid density-Wigner representation as in ref.~(14). For the projected density matrix only the real part is shown, as no element has an imaginary part larger than $1\%$ in magnitude. Unlike other states, the output with partial BSM has a modified color scheme for the projected density matrix: instead of having an intensity proportional to the magnitude, the colors represent variations relative to the output with full BSM.}
	\label{figoutputcomparison}
\end{figure}

\section*{APPENDIX D: ANALYSIS OF THE OUTPUT SWAPPED STATE - PARTIAL BSM AND DATA SET}

In the following, we discuss the effect on the output state of a partial Bell measurement and how the state reconstruction is affected by the availability of a smaller or larger tomographic data set. 

First, we specify what we mean by a ``partial'' Bell-state measurement. Our BSM consists of two parts: photon subtraction (heralded by the SNSPD) and quadrature conditioning (resolved by homodyne detection). We consider a partial BSM to be a measurement where we herald photon subtraction but do not condition on quadrature. There is of course another possibility: bypassing the photon subtraction and exercising only quadrature conditioning. This scenario is however ill-defined, as the idea of using homodyne detection to filter for specific quadrature values is specialized for the state expected after photon subtraction.

In Fig.~\ref{figoutputcomparison} we compare the output state when the BSM is performed fully (as in the main text), partially, or not at all. Since we perform our quadrature conditioning offline, we can study the case of a partial BSM (Fig.~\ref{figoutputcomparison}(b)) based on the same data as the full results (Fig.~\ref{figoutputcomparison}(a)). The incomplete realization of the BSM has a major consequence: as expected, the state bears a weakened resemblance to the targeted entangled state compared to when the BSM is completed. The projected density matrix of Fig.~\ref{figoutputcomparison}(b) highlights the variations observed between the two cases, with red strides indicating a decrease and green strides an increase when the BSM includes quadrature conditioning. Even though some coherence is already visible after the single photon-subtraction operation, the completion of the BSM depletes the vacuum component to strengthen the entanglement.

Let us note that, without conditioning, more points remain available for the tomographic reconstruction of the state -- approximately $\approx26500$, instead of only $7800$. This renders the Wigner representation of the state visibly smoother, although a much larger data set is needed to match the quality of the other states. It is worth remarking that this concern should not affect the evaluation of the state. We can for example consider the output data without any photon subtraction or conditioning, i.e., without performing a BSM at all. In this case the count rate is not as limited and we have access to a much larger data set, with up to $200000$ samples (Fig.~\ref{figoutputcomparison}(c)). Taking a subset of only $7800$ samples (Fig.~\ref{figoutputcomparison}(d)), we can directly compare with the main result to see that indeed the smoothness is lost. However, no significant coherence terms can be observed in either the Wigner representation or the projected density matrix. We conclude therefore that the relative error in the reconstruction is sufficiently small and the entanglement observed is not a product of artefacts due to limited sampling.

\section*{APPENDIX E: ESTIMATING THE NEGATIVITY OF ENTANGLEMENT}

The elements of a density matrix reconstructed tomographically are subject to higher uncertainties when a limited number of samples is available. The resulting quantum state may be only partially affected as long as the absolute error is still small relative to the value of the main contributing elements. However, as explained by Takahashi \textit{et al.}~(41), this may become a problem when trying to estimate the entanglement of the state, as the high-occupancy elements expected to have near-zero contributions may acquire finite values that statistically add up to an overestimation of the entanglement negativity.

Instead of directly measuring the negativity of the state, one can empirically observe that the logarithmic negativity $E_\mathcal{N}$, related to the entanglement negativity $\mathcal{N}$ as $E_\mathcal{N}=\log_2(2\mathcal{N}+1)$, scales in inverse proportion to the number of samples $N$ as $E_\mathcal{N}(N)\approx E_\infty+C/\sqrt{N}$. The coefficient $E_\infty$ corresponds to the asymptotic value attainable from a data sample of infinite size. By calculating the logarithmic negativity from data sets of different size, one can therefore fit for the value of $E_\infty$ and get a better evaluation of the actual negativity. The constant $C$ depends on the size of the Hilbert space used for the reconstruction of the state. A larger space is more sensitive to sampling-related complications, incrementing the rate at which a lower number of samples will cause the negativity to be overestimated.

\begin{figure}[b!]
\begin{minipage}[b]{0.35\linewidth}
\centering
\caption{\textbf{Logarithmic negativity fits.} The logarithmic negativity of the reconstructed states is shown as a function of the number of tomographic samples, $N$. The smaller dots correspond to the logarithmic negativity for each individual partition. The larger dots represent the average over partitions of the same size, with the leftmost points corresponding to the full data sets. In the case of the swapped output state, the dark, small green dots on the right correspond to the logarithmic negativities from each original run, before pooling the data together for partitioning.}
\label{fignegativityestimation}
\vspace{0cm}
\end{minipage}
\begin{minipage}[b]{0.6\linewidth}
\includegraphics[width=1.1\columnwidth]{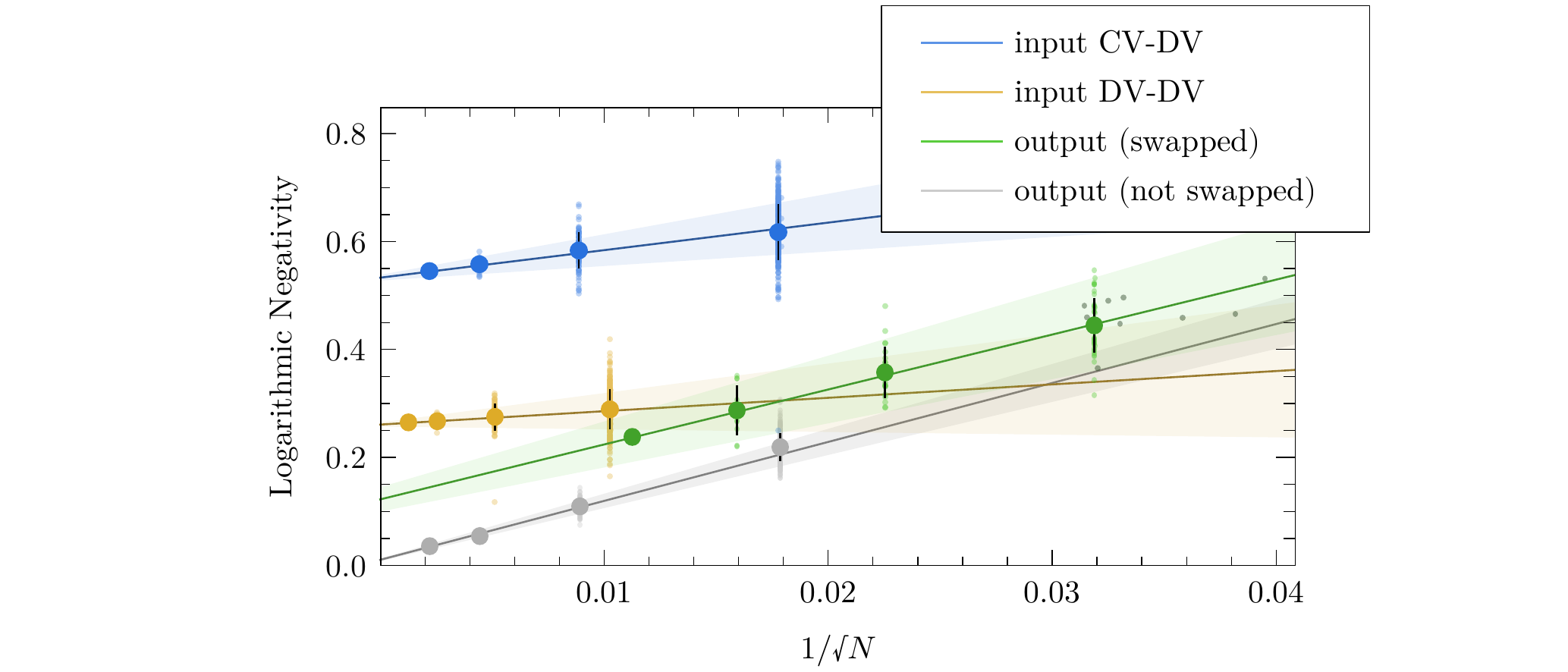}
\end{minipage}
\end{figure}

In Fig.~\ref{fignegativityestimation} we show the logarithmic negativity of the states seen in the experiment: the discrete-variable (yellow) and hybrid (blue) entangled input states, the swapped entanglement (green) and also the output state when no Bell-state measurement is performed (gray). We subdivide our data sets into consecutively smaller partitions. For a better sampling of statistical variations each partition is repeated six times, each time using a different randomly shuffled configuration. We then calculate the average logarithmic negativity for partitions of the same size and use these to fit for $E_\infty$, obtaining the following values: $0.261\pm0.004$ for the DV-DV input, $0.533\pm0.005$ for the hybrid CV-DV input, $0.010\pm0.002$ for the output without BSM and $0.122\pm0.023$ for the output when the BSM performs the entanglement swapping. Converting these values back to the typical negativity of entanglement $\mathcal{N}$, we obtain the results reported in the main text.

\section*{APPENDIX F: EFFECT OF CHANNEL LOSSES}
In this section, we detail the simulations presented in the main text. We first show the model for the ideal input states leading to the plots of the inset of Fig.~4, and then for the input experimental states.  

\subsection{Starting from ideal input states}
We first consider the effect of channel losses before the application of the BSM. Without losses, the input states can be expressed as:
\begin{equation}
\begin{aligned}
\ket{\Phi_1}_{\textrm{AB}}&=\ket{0}_{\textrm{A}}\ket{1}_{\textrm{B}}+\ket{1}_{\textrm{A}}\ket{0}_{\textrm{B}}\\
\ket{\Phi_2}_{\textrm{CD}}&=\ket{0}_{\textrm{C}}\ket{\textrm{cat}_-}_{\textrm{D}}+\ket{1}_{\textrm{C}}\ket{\textrm{cat}_+}_{\textrm{D}}.
\end{aligned}
\end{equation}
The channel losses $\eta$ are then introduced by adding virtual beamsplitter operators of transmission $\eta_{\textrm{B}}$ and $\eta_{\textrm{C}}$ on modes B and C, such that $\eta_{\textrm{B}}\eta_{\textrm{C}}=\eta$. The beamsplitters couple modes B and C with environment modes respectively denoted E and F. After introducing losses and mixing modes B and C with perfect balancing, we obtain the expression of the lossy state at the BSM station corresponding to Eq.~3 of the main text:
\begin{equation}
\begin{aligned}
\ket{\psi_{\textrm{BSM}}^{\eta}}\propto&\sqrt{\eta_{\textrm{B}}}\ket{0}_{\textrm{B}}\ket{1}_{\textrm{C}}\ket{0}_{\textrm{A}}\ket{\textrm{cat}_{-}}_{\textrm{D}}\ket{0}_{\textrm{E}}\ket{0}_{\textrm{F}}+\sqrt{\eta_{\textrm{C}}}\ket{0}_{\textrm{B}}\ket{1}_{\textrm{C}}\ket{1}_{\textrm{A}}\ket{\textrm{cat}_{+}}_{\textrm{D}}\ket{0}_{\textrm{E}}\ket{0}_{\textrm{F}}+\\
&\sqrt{\eta_{\textrm{B}}\eta_{\textrm{C}}}\ket{0}_{\textrm{B}}\ket{2}_{\textrm{C}}\ket{0}_{\textrm{A}}\ket{\textrm{cat}_{+}}_{\textrm{D}}\ket{0}_{\textrm{E}}\ket{0}_{\textrm{F}}+\\
&\sqrt{\eta_{\textrm{B}}(1-\eta_{\textrm{C}})}\ket{0}_{\textrm{B}}\ket{1}_{\textrm{C}}\ket{0}_{\textrm{A}}\ket{\textrm{cat}_{+}}_{\textrm{D}}\ket{0}_{\textrm{E}}\ket{1}_{\textrm{F}}+\\
&\sqrt{\eta_{\textrm{C}}(1-\eta_{\textrm{B}})}\ket{0}_{\textrm{B}}\ket{1}_{\textrm{C}}\ket{0}_{\textrm{A}}\ket{\textrm{cat}_{+}}_{\textrm{D}}\ket{1}_{\textrm{E}}\ket{0}_{\textrm{F}},\\
\end{aligned}
\end{equation}   
where we have kept only terms with at least one photon in mode C.

Now we apply the experimental approximation of the $\ket{1}\bra{1}_{\textrm{C}}$ operator to $\ket{\psi_{\textrm{BSM}}^{\eta}}\bra{\psi_{\textrm{BSM}}^{\eta}}$, starting from the detection. We introduce a beamsplitter of reflection $R$ to mix mode C with a new mode G, then project onto a bucket detection by applying the operator $\sum\limits_{n\geq 1}\ket{n}\bra{n}_{G}$, and finally trace out over modes B,E,F and G to obtain
\begin{equation}
\begin{aligned}
\rho_{\textrm{BSM}}^{\eta,\textrm{det}}\propto &\eta_{\textrm{B}}\ket{0}\bra{0}_{\textrm{C}}\ket{0,\textrm{cat}_-}\bra{0,\textrm{cat}_-}_{\textrm{AD}}+\eta_{\textrm{C}}\ket{0}\bra{0}_{\textrm{C}}\ket{1,\textrm{cat}_+}\bra{1,\textrm{cat}_+}_{\textrm{AD}}+\\
&\sqrt{\eta_{\textrm{B}}\eta_{\textrm{C}}}\ket{0}\bra{0}_{\textrm{C}}(\ket{1,\textrm{cat}_+}\bra{0,\textrm{cat}_-}_{\textrm{AD}}+\ket{0,\textrm{cat}_-}\bra{1,\textrm{cat}_+}_{\textrm{AD}})+\\
&(\eta_{\textrm{B}}\eta_{\textrm{C}}R+(\eta_{\textrm{B}}+\eta_{\textrm{C}}-2\eta_{\textrm{B}}\eta_{\textrm{C}}))\ket{0}\bra{0}_{\textrm{C}}\ket{0,\textrm{cat}_+}\bra{0,\textrm{cat}_+}_{\textrm{AD}}+\\
&2\eta_{\textrm{B}}\eta_{\textrm{C}}(1-R)\ket{1}\bra{1}_{\textrm{C}}\ket{0,\textrm{cat}_+}\bra{0,\textrm{cat}_+}_{\textrm{AD}}+\\
&\eta_{\textrm{B}}\sqrt{2\eta_{\textrm{C}}(1-R)}(\ket{0}\bra{1}_{\textrm{C}}\ket{0,\textrm{cat}_-}\bra{1,\textrm{cat}_+}_{\textrm{AD}}+\ket{1}\bra{0}_{\textrm{C}}\ket{1,\textrm{cat}_+}\bra{0,\textrm{cat}_-}_{\textrm{AD}})+\\
&\eta_{\textrm{C}}\sqrt{2\eta_{\textrm{B}}(1-R)}(\ket{0}\bra{1}_{\textrm{C}}\ket{0\textrm{cat}_-}\bra{1,\textrm{cat}_+}_{\textrm{AD}}+\ket{1}\bra{0}_{\textrm{C}}\ket{1,\textrm{cat}_+}\bra{0,\textrm{cat}_-}_{\textrm{AD}}).
\end{aligned}
\end{equation}  
After the application of homodyne conditioning, corresponding to the transformation $\rho\longrightarrow \rho'\propto\textrm{Tr}_{\textrm{C}}(\int_{-\Delta/2}^{+\Delta/2}dx\ket{x}\bra{x}_{\textrm{C}}\rho)$, we obtain the final expression of the swapped state:
\begin{equation}
\begin{aligned}
\rho_{\textrm{swap}}\propto &A^{\Delta}_{00}(\eta_{\textrm{B}}\ket{0,\textrm{cat}_-}\bra{0,\textrm{cat}_-}_{\textrm{AD}}+\eta_{\textrm{C}}\ket{1,\textrm{cat}_+}\bra{1,\textrm{cat}_+}_{\textrm{AD}})+\\
&A^{\Delta}_{00}\sqrt{\eta_{\textrm{B}}\eta_{\textrm{C}}}(\ket{1,\textrm{cat}_+}\bra{0,\textrm{cat}_-}_{\textrm{AD}}+\ket{0,\textrm{cat}_-}\bra{1,\textrm{cat}_+}_{\textrm{AD}})+\\
&A^{\Delta}_{00}(\eta_{\textrm{B}}\eta_{\textrm{C}}R+(\eta_{\textrm{B}}+\eta_{\textrm{C}}-2\eta_{\textrm{B}}\eta_{\textrm{C}}))\ket{0,\textrm{cat}_+}\bra{0,\textrm{cat}_+}_{\textrm{AD}}+\\
&A^{\Delta}_{11}2\eta_{\textrm{B}}\eta_{\textrm{C}}(1-R)\ket{0,\textrm{cat}_+}\bra{0,\textrm{cat}_+}_{\textrm{AD}},
\end{aligned}
\label{swapIdeal}
\end{equation}
where $A^{\Delta}_{ij}=\textrm{Tr}_{\textrm{C}}(\int_{-\Delta/2}^{\Delta/2}dx\ket{x}\bra{x}_{\textrm{C}}\ket{i}\bra{j}_{\textrm{C}})$, and noticing that $A^{\Delta}_{ij}=0$ if $i-j=\pm 1$. The first four terms correspond to the ideal maximally entangled swapped state for $\eta_{\textrm{B}}=\eta_{\textrm{C}}=1$ and have to be favored to obtain the highest negativity after swapping. Conversely, the last two terms correspond to pure losses and should be minimized for optimal results. Fixing $\eta=\eta_{{\textrm{B}}}\eta_{{\textrm{C}}}$, one notices that the second-to-last term is minimum for $\eta_{{\textrm{C}}}=\sqrt{\eta}$, \textit{i.e.} for symmetric losses. The plots of Fig.~4 are therefore all drawn in the case of symmetric losses $\eta_{{\textrm{B}}}=\eta_{{\textrm{C}}}$. 
The final results are found by adding to the model losses $(1-\eta_{\textrm{HD}})$ on the homodyne detection path. This simply translates to applying to Eq.~\ref{swapIdeal} the transformation $A^{\Delta}_{11}\rightarrow \eta_{\textrm{HD}}A^{\Delta}_{11}+(1-\eta_{\textrm{HD}})A^{\Delta}_{00}$.\\
As losses increase, the probability of observing a click due to the presence of two input photons before mixing on the BSM beamsplitter becomes a dominant term in the losses. Indeed the second-to-last term in Eq.~\ref{swapIdeal} is equivalent at greater losses to:
\begin{equation}
2A^{\Delta}_{00}\sqrt{\eta}\ket{0,\textrm{cat}_+}\bra{0,\textrm{cat}_+}_{\textrm{AD}}.
\end{equation}
leading to the expression of the swapped state over an infinitely lossy channel:
\begin{equation}
\begin{aligned}
\rho_{\textrm{swap}}=0.5\ket{\Phi_2}\bra{\Phi_2}_{\textrm{AD}}+0.5\ket{0,\textrm{cat}_+}\bra{0,\textrm{cat}_+}_{\textrm{AD}}.
\end{aligned}
\end{equation}
This state exhibits an entanglement negativity of $\mathcal{N}\approx 0.104$ and is the asymptotic limit of the negativity after swapping at high losses starting from maximally entangled states. It also corresponds to a swapping protocol with ideal photon subtraction (vanishing R) but no homodyne conditioning. This derivation in the ideal case is similar and in agreement with previous studies of DV entanglement swapping in a lossy channel (42). In the following, we adapt the reasoning to the case of non maximally-entangled input states.

\subsection{Using the experimental states as input}
The model we choose for realistic input states is a mixture of these two states with fully mixed classical superpositions and with vacuum: 

\begin{equation}
\begin{aligned}
\rho_{1}&\propto cg_1\ket{\Phi_1}\bra{\Phi_1}+0.5cm_1(\ket{0,1}\bra{0,1}+\ket{1,0}\bra{1,0})+cv_1\ket{0,0}\bra{0,0} \\
\rho_{2}&\propto cg_2\ket{\Phi_2}\bra{\Phi_2}+0.5cm_2(\ket{0,\textrm{cat}_-}\bra{0,\textrm{cat}_-}+\ket{1,\textrm{cat}_+}\bra{1,\textrm{cat}_+})+cv_2\ket{0,\textrm{cat}_+}\bra{0,\textrm{cat}_+}.
\end{aligned}
\end{equation}
We find the best agreement with the experiment's measured input states using the parameters:
\begin{center}
\begin{tabular}{ll}
$cg_1=1$ & $cg_2=1$\\
$cm_1=0.05$ & $cm_2=0.047$\\
$cv_1=0.97$ & $cv_2=0.438$.
\end{tabular}
\end{center}
Following the same reasoning as in section~A we then compute the expression of:
\begin{equation}
\rho_{\textrm{swap}}^{\textrm{exp},0}=\textrm{Tr}_{\textrm{BCEFG}}(\int_{-\Delta/2}^{+\Delta/2}dx\ket{x}\bra{x}_{\textrm{C}}\sum\limits_{n\geq 1}\ket{n}\bra{n}_{\textrm{G}}B^{1-R}_{\textrm{CG}}B_{\textrm{BC}}^{\frac{1}{2}}{B}^{\eta_{\textrm{B}}}_{\textrm{BE}}{B}^{\eta_{\textrm{C}}}_{\textrm{CF}}\rho_1\otimes\rho_2),
\end{equation}
where ${B}^{T}_{\textrm{XY}}$ is the beamsplitter operator of transmission $T$ mixing modes X and Y. To that expression we have to add as well the effect of false positive events and dark counts. False positive events are due to the imperfect extinction from the AOM switch used to block the first heralding event as well as the ratio between the heralding rate of input states and of BSM detection events. The integration of the histograms of Fig.~\ref{figtimefilteringI} over the 8-ns time-window considered leads us to estimate the false positive events to represent close to $40\%$ of the total events recorded. This leads to a refinement of the model to obtain:
\begin{equation}
\rho_{\textrm{swap}}^{\textrm{exp,FP}}=0.6\rho_{\textrm{swap}}^{\textrm{exp},0}+0.4\ket{0,\textrm{cat}_+}\bra{0,\textrm{cat}_+}.
\end{equation}

Note that this issue could be strongly minimized through the addition of a fast shutter or an AOM at the output of the type-II OPO to lower the probability of coincidental down-conversion events and therefore the noise-floor of Fig.~\ref{figtimefilteringI}. Using an AOM would lead to a reduction by a factor five of false positive events, lowering their proportion from $40\%$ to $8\%$, although at the cost of a significant loss of efficiency for the total protocol. The use of a mechanical shutter would aleviate this problem and could be used in a future implementation of the experiment, although it entails a challenging extension of the delay line. Finally, as the effect is stronger because of the low rate of generation of the input states, the use of three OPOs instead of two would greatly mitigate this problem. Ultimately, this effect is completely dependent on our current setup and is not a fundamental issue of the protocol for the BSM. An effect of broader interest however is that of dark counts at the detector level. Its effect is dependent on the channel as the relative number of dark counts over positive events will increase with higher losses. At zero losses, we measure the ratio, noted $\eta_d$, to be one dark count for one hundred positive events. The final expression of the swapped experimental state with dark counts is:
\begin{equation}
\rho_{\textrm{swap}}^{\textrm{exp,FP,dark}}=\rho_{\textrm{swap}}^{\textrm{exp,FP}}+\frac{\eta_d}{\eta_{\textrm{B}}\eta_{\textrm{C}}}\ket{0,\textrm{cat}_+}\bra{0,\textrm{cat}_+}.
\end{equation}

\end{document}